\documentclass[
reprint, 
superscriptaddress,
showpacs,
preprintnumbers,
amsmath,
amssymb,
aps,
prl,
]{revtex4-1}

\usepackage{appendix}
\usepackage{siunitx}
\usepackage{graphicx}   
\usepackage{hyperref}
\usepackage{booktabs}
\usepackage{braket}
\usepackage[colorinlistoftodos,prependcaption,textsize=tiny]{todonotes}
\usepackage{changes}
\usepackage[version=4]{mhchem}
\usepackage{upgreek}

\DeclareGraphicsExtensions{.pdf, .png, .jpg, .jpeg, .svg}

\begin{document}


\title{Magnons at low excitations: Observation of incoherent coupling to a bath of two-level systems}

\author{Marco Pfirrmann}
	\email{marco.pfirrmann@kit.edu}
	\affiliation{Institute of Physics, Karlsruhe Institute of Technology, 76131 Karlsruhe, Germany}
\author{Isabella Boventer}
	\affiliation{Institute of Physics, Karlsruhe Institute of Technology, 76131 Karlsruhe, Germany}
	\affiliation{Institute of Physics, Johannes Gutenberg-University Mainz, 55099 Mainz, Germany}
\author{Andre Schneider}
	\affiliation{Institute of Physics, Karlsruhe Institute of Technology, 76131 Karlsruhe, Germany}	\author{Tim Wolz}
	\affiliation{Institute of Physics, Karlsruhe Institute of Technology, 76131 Karlsruhe, Germany}\author{Mathias Kl\"aui}
	\affiliation{Institute of Physics, Johannes Gutenberg-University Mainz, 55099 Mainz, Germany}
\author{Alexey V. Ustinov}
	\affiliation{Institute of Physics, Karlsruhe Institute of Technology, 76131 Karlsruhe, Germany}
	\affiliation{Russian Quantum Center, National University of Science and Technology MISIS, 119049 Moscow, Russia}
\author{Martin Weides}
	\email{martin.weides@glasgow.ac.uk}
	\affiliation{Institute of Physics, Karlsruhe Institute of Technology, 76131 Karlsruhe, Germany}
	\affiliation{James Watt School of Engineering, University of Glasgow, Glasgow G12 8LT, United Kingdom}

\date{\today}

\begin{abstract}
Collective magnetic excitation modes, magnons, can be coherently coupled to microwave photons in the single excitation limit. This allows for access to quantum properties of magnons and opens up a range of applications in quantum information processing, with the intrinsic magnon linewidth representing the coherence time of a quantum resonator. Our measurement system consists of a yttrium iron garnet sphere and a three-dimensional microwave cavity at temperatures and excitation powers typical for superconducting quantum circuit experiments. We perform spectroscopic measurements to determine the limiting factor of magnon coherence at these experimental conditions. Using the input-output formalism, we extract the magnon linewidth $\kappa_\mathrm{m}$. We attribute the limitations of the coherence time at lowest temperatures and excitation powers to incoherent losses into a bath of near-resonance two-level systems (TLSs), a generic loss mechanism known from superconducting circuits under these experimental conditions. We find that the TLSs saturate when increasing the excitation power from quantum excitation to multiphoton excitation and their contribution to the linewidth vanishes. At higher temperatures, the TLSs saturate thermally and the magnon linewidth decreases as well.
\end{abstract}

\maketitle

Strongly coupled light-spin hybrid systems allow for coherent exchange of quantum information. Such systems are usually studied either classically at room temperature \cite{Zhang_2014} or at millikelvin temperatures approaching the quantum limit of excitation \cite{Huebl_2013, Tabuchi_2014, Goryachev_2014}. The field of cavity magnonics \cite{Zhang_2015, Bai_2015, Cao_2015, Goryachev_PRB_2018, Boventer_2018} harnesses the coherent exchange of excitation due to the strong coupling within the system and is used to access a new range of applications such as quantum transducers and memories \cite{Zhang_NatComm_2015}.
Nonlinearity in the system is needed to gain access to the control and detection of single magnons. Because of experimental constraints regarding required light intensities in a purely optomagnonic system \cite{Kusminskiy_2016}, hybridized systems of magnon excitations and non-linear macroscopic quantum systems such as superconducting qubits \cite{Tabuchi_2015, Tabuchi_2016} are used instead, which opens up new possibilities in the emerging field of quantum magnonics \cite{Lachance-Quirion_2017, Lachance-Quirion_2019}. An efficient interaction of magnonic systems and qubits requires their lifetimes to exceed the exchange time. Magnon excitation losses, expressed by the magnon linewidth $\kappa_\mathrm{m}$, translate into a lifetime of the spin excitation. Identifying its limiting factors is an important step toward more sophisticated implementations of hybrid quantum systems using magnons. Studies in literature show the losses in magnon excitations from room temperature down to about liquid helium temperatures \cite{Boventer_2018}. The main contribution changes with temperature from scattering at rare-earth impurities \cite{Spencer_1961, Seiden_1964} to multi-magnon scattering at imperfect sample surfaces \cite{Sparks_1961, Nemarich_1964}. For a typical environment of superconducting quantum circuit experiments, temperatures below $\SI{100}{\milli\kelvin}$ and microwave probe powers comparable to single-photon excitations, temperature sweeps show losses into TLSs \cite{Tabuchi_2014}. In this paper, we present both temperature- and power-dependent measurements of the magnon linewidth in a spherical yttrium iron garnet (YIG) sample in the quantum limit of magnon excitations. We extract the critical saturation power and present on- and off-resonant linewidth that is mapped to the ratio of magnon excitation in the hybrid system. For large detuning, the fundamental linewidth can be extracted, thereby avoiding unwanted saturation effects from the residual cavity photon population. This renders the off-resonant linewidth a valuable information on the limiting factors of spin lifetimes.

\begin{figure*}[hbt]
\centering
\includegraphics[width=\textwidth]{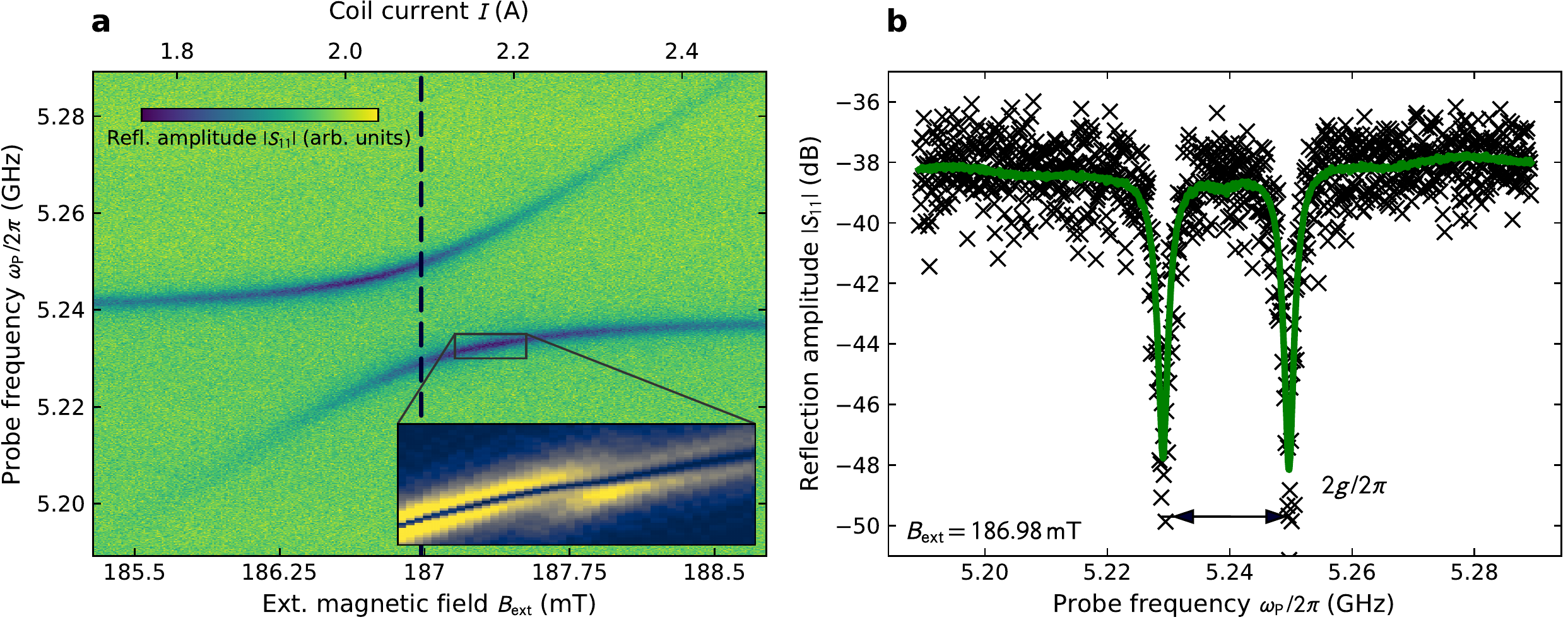}
\caption{
(a) Color coded absolute value of the reflection spectrum plotted against probe frequency and applied current at $T=\SI{55}{\milli\kelvin}$ and $P=-140\,\mathrm{dBm}$. The resonance dips show the dressed photon-magnon states forming an avoided level crossing with the degeneracy point at $I_0=\SI{2.09}{\ampere}$, corresponding to an applied field of $B_0=\SI{186.98}{\milli\tesla}$ (dashed vertical line). The inset displays the squared gradient of the zoomed-in amplitude data. The kink in the data represents a weakly coupled magnetostatic mode. This was also seen in Ref. \cite{Boventer_2018}. 
(b) Raw data of the cross section at the center of the avoided level crossing and fit to the input-output formalism. The fit gives a magnon linewidth of $\kappa_\mathrm{m}/2\uppi=\SI[separate-uncertainty, multi-part-units=single]{1.82 \pm 0.18}{\mega\hertz}$. The data are normalized by the field independent background before fitting and is multiplied to the fit to display it over the raw data.
}
\label{fig:combined_anticrossing}
\end{figure*}

The magnetization dynamics inside a magnetic crystal is described by bosonic quasiparticles of collective spin excitation, called magnons. These magnons manifest as the collective precessional motion of the participating spins out of their equilibrium positions. Their energies and spatial distribution can be calculated analytically using the Walker modes for spherical samples \cite{Walker_1957, Fletcher_1959}. We focus on the uniform in-phase precession mode corresponding to the wave vector $\mathbf{k}=0$, called the Kittel mode \cite{Kittel_1948}, treating it equivalently to one single large macro spin. The precession frequency (magnon frequency) of the Kittel mode in a sphere changes linearly with a uniform external magnetic bias field. The precessional motion is excited by a magnetic field oscillating at the magnon frequency perpendicular to the bias field. We use the confined magnetic field of a cavity photon resonance to create magnetic excitations in a macroscopic sample, biased by a static external magnetic field. Tuning them into resonance, the magnon and photon degree of freedom mix due to their strong interaction. This creates hybridized states described as repulsive cavity magnon polaritons, which are visible as an avoided level crossing in the spectroscopic data with two resonance dips at frequencies $\omega_\pm$ (see Supplemental Material \cite{supp}) appearing in the data cross section. The interaction is described by the macroscopic magnon-photon coupling strength $g$. The system is probed in reflection with microwave frequencies using standard ferromagnetic resonance techniques \cite{Kalarickal_2006}. We use the input-output formalism \cite{QuantumOpticsBook} to describe the reflection spectrum. The complex reflection parameter $\mathcal{S}_{11}$, the ratio of reflected to input energy with respect to the probe frequency $\omega_\mathrm{p}$, reads as
\begin{equation}
	\mathcal{S}_{11}(\omega_\mathrm{p})=-1 +\frac{2\kappa_\mathrm{c}} {\mathrm{i} \left(\omega_\mathrm{r}-\omega_\mathrm{p}\right) +\kappa_\mathrm{l}+ \frac{g^2}{\mathrm{i}\left(\omega_\mathrm{m}-\omega_\mathrm{p}\right)+\kappa_\mathrm{m}}},
	\label{eq:S11}
\end{equation}
with the cavity's coupling and loaded linewidths $\kappa_\mathrm{c}$ and $\kappa_\mathrm{l}$, and the internal magnon linewidth $\kappa_\mathrm{m}$ (HWHM).
\begin{figure*}[htb]
\centering
\includegraphics[width=\textwidth]{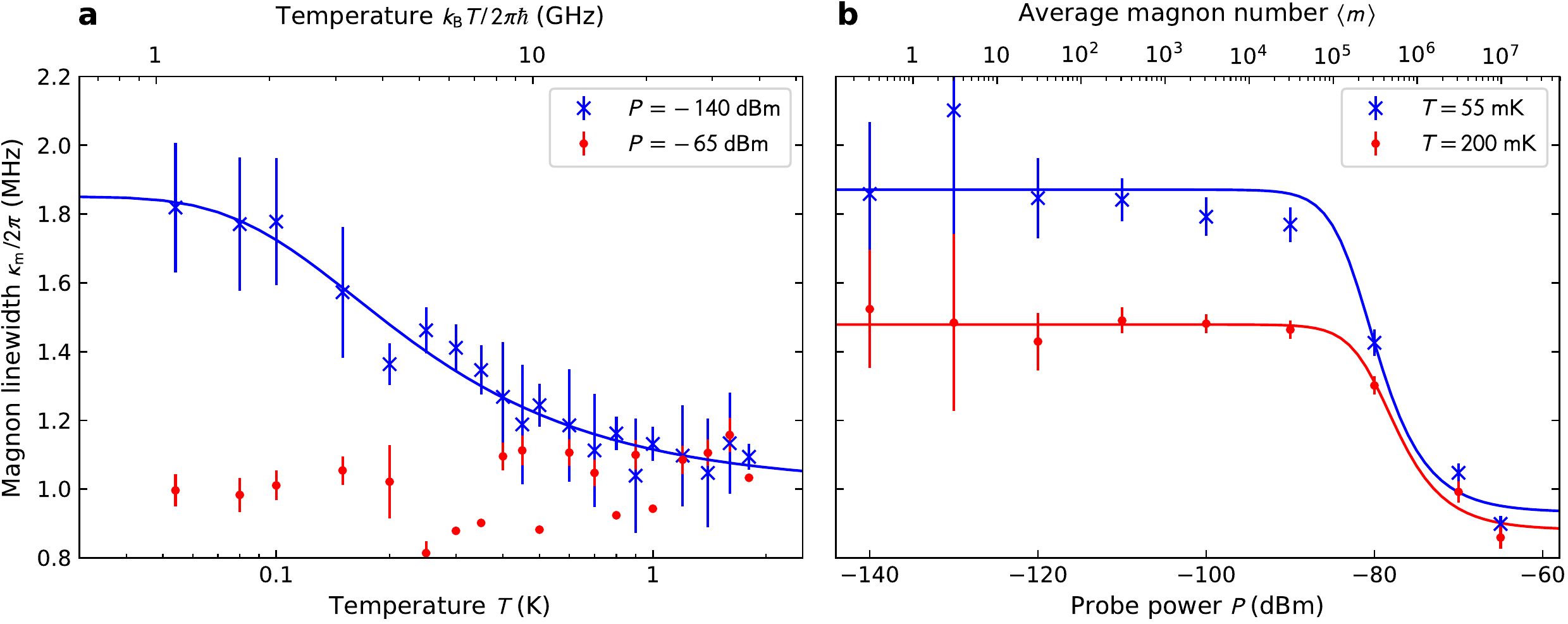}
\caption{
(a) Temperature dependence of the magnon linewidth $\kappa_\mathrm{m}$ at the degeneracy point. For low probe powers, $\kappa_\mathrm{m}$ follows a $\tanh\left(1/T\right)$ behavior (crosses), while for high probe powers (circles) the linewidth does not show any temperature dependence.
(b) Power dependence of the magnon linewidth $\kappa_\mathrm{m}$ for $T=\SI{55}{\milli\kelvin}$ and $\SI{200}{\milli\kelvin}$ at the degeneracy point. Both temperature curves show a similar behavior. At probe powers of about $\SI{-90}{\deci\bel\metre}$ $\kappa_\mathrm{m}$ drops for both temperatures, following the $\left(1+P/P_\mathrm{c} \right)^{-1/2}$ trend of the TLS model. All linewidth data shown here are extracted from the fit at matching frequencies.
}
\label{fig:linewidth_combined}
\end{figure*}

For our hybrid system we mount a commercially available YIG (\ce{Y3Fe5O12}) sphere with a diameter $d=\SI{0.5}{\milli\metre}$ \cite{ferrisphere} inside a three-dimensional (3D) rectangular cavity made of oxygen-free copper and cool the device in a dilution refrigerator down to millikelvin temperatures (see figure in Supplemental Material \cite{supp}). YIG as a material is particularly apt for microwave applications, as it is a ferrimagnetic insulator with a very low Gilbert damping factor of $10^{-3}$ to $10^{-5}$  \cite{Kajiwara_2010, Heinrich_2011, Kurebayashi_2011} and a high net spin density of $2.1\times 10^{22}\,\si[per-mode=symbol]{\mu_\mathrm{B}\per\centi\cubic\metre}$ \cite{Gilleo_1958}. The single crystal sphere comes pre-mounted to a beryllium oxide rod along the [110] crystal direction.
The 3D cavity has a $\mathrm{TE_{102}}$ mode resonance frequency of $\omega_\mathrm{r}^\mathrm{bare}/2\uppi = \SI{5.24}{\giga\hertz}$ and is equipped with one SMA connector for reflection spectroscopy measurements. For low temperatures and excitation powers, we find the internal and coupling quality factors to be $Q_\mathrm{i}=\omega_\mathrm{r}/2\kappa_\mathrm{i}=7125\pm 97$ and $Q_\mathrm{c}=\omega_\mathrm{r}/2\kappa_\mathrm{c}=5439\pm 29$, combining to a loaded quality factor $Q_\mathrm{l}=\left(1/Q_\mathrm{i}+1/Q_\mathrm{c}\right)^{-1}=3084\pm 24$ (see Supplemental Material \cite{supp}). 
We mount the YIG at a magnetic anti-node of the cavity resonance and apply a static magnetic field  of about $\SI{187}{\milli\tesla}$ perpendicular to the cavity field to tune the magnetic excitation into resonance with the cavity photon. The magnetic field is created by an iron yoke holding a superconducting niobium-titanium coil. Additional permanent samarium-cobalt magnets are used to create a zero-current offset magnetic field of about $\SI{178}{\milli\tesla}$.
The probing microwave signal is provided by a vector network analyzer (VNA). Microwave attenuators and cable losses account for $\SI{-75}{\deci\bel}$ of cable attenuation to the sample. We apply probe powers between $-140\,\mathrm{dBm}$ and $-65\,\mathrm{dBm}$ at the sample's SMA port. Together with the cavity parameters, this corresponds to an average magnon population number $\left<m\right>$ from $0.3$ up to the order of $10^7$ \cite{supp} in the hybridized case. The probe signal is coupled capacitively to the cavity photon using the bare inner conductor of a coaxial cable positioned in parallel to the electric field component.
The temperature of the sample is swept between $\SI{55}{\milli\kelvin}$ and $\SI{1.8}{\kelvin}$ using a proportional-integral-derivative (PID) controlled heater. After a change in temperature, we wait at least one hour for the sample to thermalize before measuring. All data acquisition and analysis are done via qkit \cite{qkit}.

A typical measurement is shown in Fig.\@ \ref{fig:combined_anticrossing}(a), measured at $T=\SI{55}{\milli\kelvin}$ with an input power level of $P=-140\,\mathrm{dBm}$. Figure\@ \ref{fig:combined_anticrossing}(b) shows the raw data and the fit of the cavity-magnon polariton at matching resonance frequencies for an applied external field of $B_0=\SI{186.98}{\milli\tesla}$. We correct the raw data from background resonances and extract the parameters of the hybridized system by fitting to Eq.\@ (\ref{eq:S11}). The coupling strength $g/2\uppi = \SI{10.4}{\mega\hertz}$ of the system exceeds both the total resonator linewidth $\kappa_\mathrm{l}/2\uppi = \frac{\omega_\mathrm{r}}{2Q_\mathrm{l}}/2\uppi=\SI{0.85}{\mega\hertz}$ and the internal magnon linewidth $\kappa_\mathrm{m}/2\uppi = \SI{1.82}{\mega\hertz}$, thus being well in the strong coupling regime ($g \gg \kappa_\mathrm{l},\,\,\kappa_\mathrm{m}$) for all temperatures and probe powers. The measured coupling strength is in good agreement with the expected value

\begin{equation}
	g^\mathrm{th}=\frac{\gamma_e\eta}{2}\sqrt{\frac{\mu_0\hbar\omega_\mathrm{r}}{2V_\mathrm{a}}}\sqrt{2N_\mathrm{s}s},
	\label{eq:g}
\end{equation}
with the gyromagnetic ratio of the electron $\gamma_e$, the mode volume $V_\mathrm{a}=\SI{5.406e-6}{\cubic\metre}$, the \ce{Fe^3+} spin number $s=5/2$, the spatial overlap between microwave field and magnon field $\eta$, and the total number of spins $N_\mathrm{s}$ \cite{Boventer_2018}. The overlap factor is given by the ratio of mode volumes in the cavity volume and the sample volume \cite{Zhang_2014}. We find for our setup the overlap factor to be $\eta = 0.536$. For a sphere diameter of $d=\SI{0.5}{\milli\metre}$ we expect a total number of $N_\mathrm{s}=1.37\times 10^{18}$ spins. We find the expected coupling strength $g^\mathrm{th}/2\uppi=\SI{12.48}{\mega\hertz}$ to be in good agreement with our measured value.
Even for measurements at high powers, the number of participating spins of the order of $10^{18}$ is much larger than the estimated number of magnon excitations ($\sim 10^{7}$). We therefore do not expect to see the intrinsic magnon nonlinearity as observed at excitation powers comparable to the number of participating spins \cite{Haigh_2015}.
\begin{figure*}[htb]
\centering
\includegraphics[width=\textwidth]{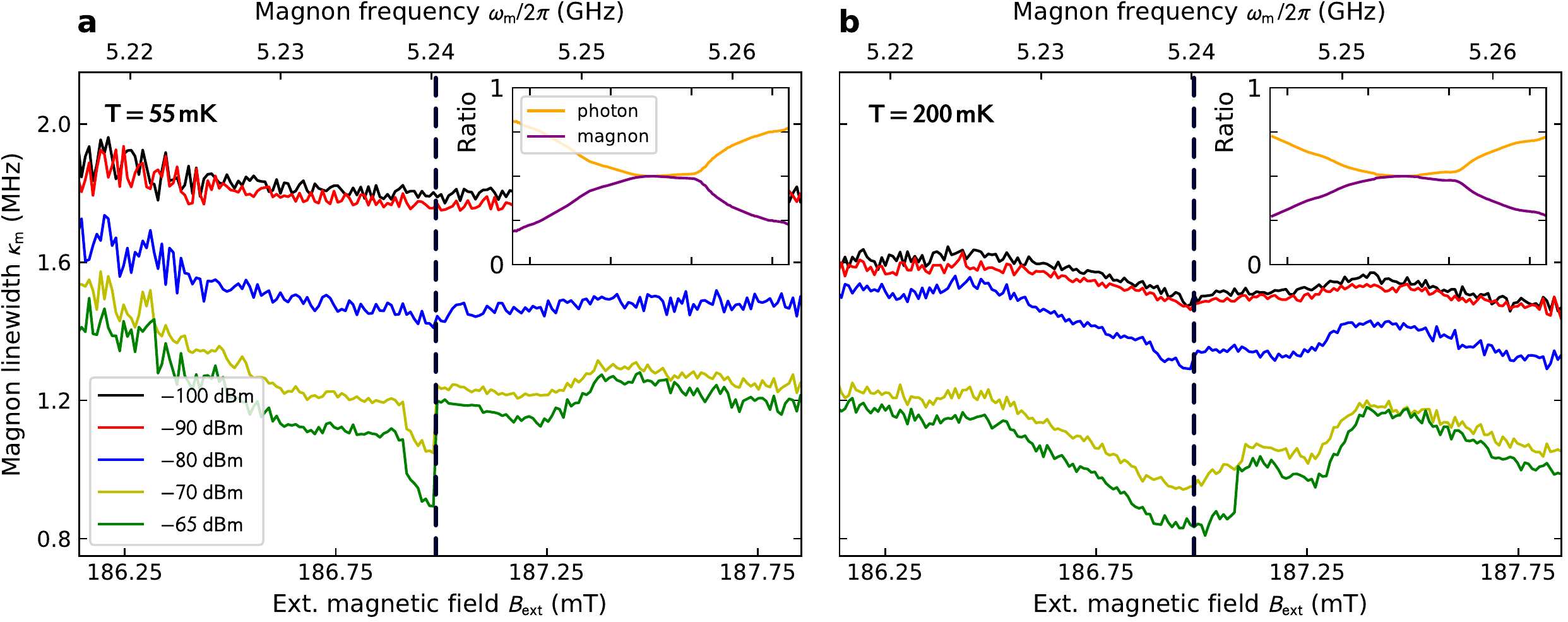}
\caption{
Magnetic field (magnon frequency) dependence of the the magnon linewidth $\kappa_\mathrm{m}$ for different probe powers at $T=\SI{55}{\milli\kelvin}$  (a) and $T=\SI{200}{\milli\kelvin}$ (b). The shown probe powers correspond to the ones at the transition in Fig.\@ \ref{fig:linewidth_combined} (b). The number of excited magnons depends on the detuning of magnon and photon frequency. At matching frequencies (dashed line) the magnon linewidth has a minimum, corresponding to the highest excited magnon numbers and therefore the highest saturation of TLSs. A second minimum at about $\SI{187.25}{\milli\tesla}$ corresponds to the coupling to an additional magnetostatic mode within the sample [inset of Fig.\@ \ref{fig:combined_anticrossing} (a)].  The insets show the ratio of excitation power within each component of the hybrid system. At matching frequencies, both components are excited equally. The magnon share drops at the plot boundaries to about $\SI{20}{\percent}$. The coupling to the magnetostatic mode is visible as a local maximum in the magnon excitation ratio. The $x$ axes are scaled as in the main plots. The legends are valid for both temperatures.
}
\label{fig:linewidth_power}
\end{figure*}

The internal magnon linewidth decreases at higher temperature and powers (Fig.\@ \ref{fig:linewidth_combined}) while the coupling strength remains geometrically determined and does not change with either temperature or power. This behavior can be explained by an incoherent coupling to a bath of two-level systems (TLSs) as the main source of loss in our measurements. In the TLS model \cite{Anderson_1972, Phillips_1972, Hunklinger_1976, Phillips_1987}, a quantum state is confined in a double-well potential with different ground-state energies and a barrier in-between. TLSs become thermally saturated at temperatures higher than their frequency ($T\gtrsim \hbar\omega_\mathrm{TLS}/k_\mathrm{B}$). Dynamics at low temperatures are dominated by quantum tunneling through the barrier that can be stimulated by excitations at similar energies. This resonant energy absorption shifts the equilibrium between the excitation rate and lifetime of the TLSs and their influence to the overall excitation loss vanishes. Loss into an ensemble of near-resonant TLSs is a widely known generic model for excitation losses in solids, glasses, and superconducting circuits at these experimental conditions \cite{Mueller_2017}. We fit the magnon linewidth to the generic TLS model loss tangent
\begin{equation}
	\label{eq:TLS}
	\kappa_\mathrm{m}\left(T,\,\,P\right)=\kappa_0\frac{\tanh\left(\hbar\omega_\mathrm{r}/2k_\mathrm{B}T\right)}{\sqrt{1+P/P_\mathrm{c}}}+\kappa_\mathrm{off}.
\end{equation}
Directly in the avoided level crossing we find $\kappa_0/2\uppi =\SI[separate-uncertainty, multi-part-units=single]{1.05 \pm 0.15}{\mega\hertz}$ as the low temperature limit of the linewidth describing the TLS spectrum within the sample and $\kappa_\mathrm{off}/2\uppi =\SI[separate-uncertainty, multi-part-units=single]{0.91 \pm 0.11}{\mega\hertz}$ as an offset linewidth added as a lower boundary without TLS contribution. The critical power $P_\mathrm{c}=-81 \pm 6.5\,\mathrm{dBm}$ at the SMA port describes the saturation of the TLS due to resonant power absorption, corresponding to an average critical magnon number of $\left<m_\mathrm{c}\right> = 2.4\cdot 10^5$. 
Using finite-element simulations, we map the critical excitation power to a critical AC magnetic field on the order of $B_\mathrm{c}\sim\SI[exponent-product = \cdot]{3e-10}{\tesla}$ at the position of the YIG sample.
Looking at the linewidths outside the anti-crossing at constant input power, we find a minimum at matching magnon and photon frequencies (dashed lines in Fig.\@ \ref{fig:linewidth_power}). Here, the excitation is equally distributed between photons and magnons, reaching the maximum in both magnon excitation power and TLS saturation, respectively. At detuned frequencies the ratio between magnon and photon excitation power changes, less energy excites the magnons (insets in Fig.\@ \ref{fig:linewidth_power}), and therefore less TLSs get saturated. The magnon linewidth increases with detuning, matching the low power data for large detunings. This effect is most visible at highest excitation powers. We calculate the energy ratios by fitting the resonances in each polariton branch individually and weight the stored energy with the eigenvalues of the coupling Hamiltonian \cite{Harder_2018}.
For higher powers a second minimum at about $\SI{187.25}{\milli\tesla}$ can be seen at both studied temperatures. We attribute this to the coupling to a magnetostatic mode within the YIG sample and therefore again an increased number of excited magnons [see inset of Fig.\@ \ref{fig:combined_anticrossing} (a)]. This can also be seen in the inset figures as a local magnon excitation maximum.
We attribute the TLS-independent losses $\kappa_\mathrm{off}/2\uppi=\SI[separate-uncertainty, multi-part-units=single]{0.91 \pm 0.11}{\mega\hertz}$ to multi-magnon scattering processes on the imperfect sphere surface \cite{Sparks_1961, Nemarich_1964}. As described in Ref. \cite{Boventer_2018}, we model the surface of the YIG with spherical pits with radii of $\frac{2}{3}$ of the size of the polishing material ($2/3 \times \SI{0.05}{\micro\metre}$) and estimate a contribution of about $2\uppi\cdot \SI{1}{\mega\hertz}$ that matches our data.
We attribute the slight increase in the linewidth visible in the high-power data (circles) in Fig.\@ \ref{fig:linewidth_combined} (a) to the first influence of rare-earth impurity scattering, dominating the linewidth behavior of the TLS-saturated system at higher temperatures \cite{Spencer_1959, Spencer_1961, Boventer_2018}.
In principle, loss due to TLS can also be determined indirectly by weak changes of the resonance frequency \cite{Pappas_2011, Gao_2008, Kumar_2008, Burnett_2014} while keeping the field constant. Our system, however, operates at fixed frequency and magnetic remanence within the magnetic yoke leads to uncertainties in absolute magnetic field value beyond the required accuracy.

In this work, we studied losses in a spherical YIG sample at temperatures below $\SI{2}{\kelvin}$ and excitation powers down from $10^7$ photons below a single photon. We identify incoherent coupling to a bath of two-level systems as the main source of excitation loss in our measurements. The magnon linewidth $\kappa_\mathrm{m}/2\uppi$ at the degeneracy point fits well to the generic loss tangent of the TLS model with respect to temperature and power. It decreases from about $\SI{1.8}{\mega\hertz}$ influenced by TLSs to about $\SI{1}{\mega\hertz}$ with saturated TLSs. The magnon linewidth shows a minimum at maximum magnon excitation numbers, again corresponding with TLS saturation with increasing excitation power. While TLSs are a common source of loss in superconducting circuits, their microscopic nature is still not fully understood. Possible models for TLS origin include magnetic TLSs in spin glasses \cite{Continentino_1981_ssc, Continentino_1981, Wassermann_1984, Wesenberg_2017} that manifest in crystalline samples in lower concentration, surface spins \cite{deGraaf_2017, deGraaf_2018} that influence the effective number of spins or magnon-phonon and subsequent phonon losses into TLSs \cite{Streib_2019} (see also Supplemental Material \cite{supp}. Improving the surface roughness and quality of the YIG crystal can lead to lower overall losses and lower TLS influence which can lead to longer coherence lifetimes for application in quantum magnonic devices.

\textit{Note added in proof} - Recently, a manuscript studying losses in thin film YIG that independently observed comparable results and reached similar conclusions was published by Kosen \textit{et al.} \cite{Kosen_2019}.

\begin{acknowledgments}
This work was supported by the European Research Council (ERC) under the Grant Agreement 648011 and the Deutsche Forschungsgemeinschaft (DFG) within Project INST 121384/138-1 FUGG and SFB TRR 173. We acknowledge financial support by the Helmholtz International Research School for Teratronics (M.P. and T.W.) and the Carl-Zeiss-Foundation (A.S.). A.V.U acknowledges partial support from the Ministry of Education and Science of Russian Federation in the framework of the Increase Competitiveness Program of the National University of Science and Technology MISIS (Grant No. K2-2017-081).
\end{acknowledgments}

\bibliography{Pfirrmann_magnon_TLS}
\bibliographystyle{apsrev4-1}

\appendix

\renewcommand{\thefigure}{S\arabic{figure}}
\renewcommand{\theequation}{S\arabic{equation}}
\setcounter{figure}{0}
\setcounter{equation}{0}
\section{\label{app:sec:cavity_magnon_coupling}Cavity-Magnon coupling}
The frequencies of both arms of the avoided level crossing $\omega_\pm$ are fitted to the energy eigenvalues of a $2\times 2$ matrix describing two coupled harmonic oscillators, one with constant frequency and one with a linearly changing frequency, 
\begin{equation}
	\omega_\pm=\frac{\omega_\mathrm{r}^\mathrm{bare}+\omega_\mathrm{m}^{I=0}}{2}\pm\sqrt{\left(\frac{\omega_\mathrm{r}^\mathrm{bare}-\omega_\mathrm{m}^{I=0}}{2}\right)^2+g^2}.
	\label{app:eq:bare_frequency}
\end{equation}
We use the current dependent data taken at $T=\SI{55}{\milli\kelvin}$ and $P=-140\,\mathrm{dBm}$ to obtain the bare cavity frequency $\omega_\mathrm{r}^\mathrm{bare}$,  the zero-current magnetic excitation frequency $\omega_\mathrm{m}^{I=0}$, and the coupling strength $g$.
The frequencies of the anticrossing $\omega_\pm$ were obtained by tracking the minima in the amplitude data. From the fit we obtain the bare cavity frequency $\omega_\mathrm{r}^\mathrm{bare}/2\uppi =\SI[separate-uncertainty, multi-part-units=single]{5.23902(2)}{\giga\hertz}$ and the zero-current magnetic excitation frequency $\omega_\mathrm{m}^{I=0}/2\uppi=\SI[separate-uncertainty, multi-part-units=single]{4.9817(2)}{\giga\hertz}$ due to the offset magnetic field by the permanent magnets. The magnon-cavity coupling strength stays nearly constant for all temperatures and excitation powers at $g/2\uppi=\SI[separate-uncertainty, multi-part-units=single]{10.39\pm0.17}{\mega\hertz}$.
\section{\label{app:sec:photon_number_estimation}Magnon number estimation}
\begin{figure*}[hbt]
\centering
\includegraphics[width=\textwidth]{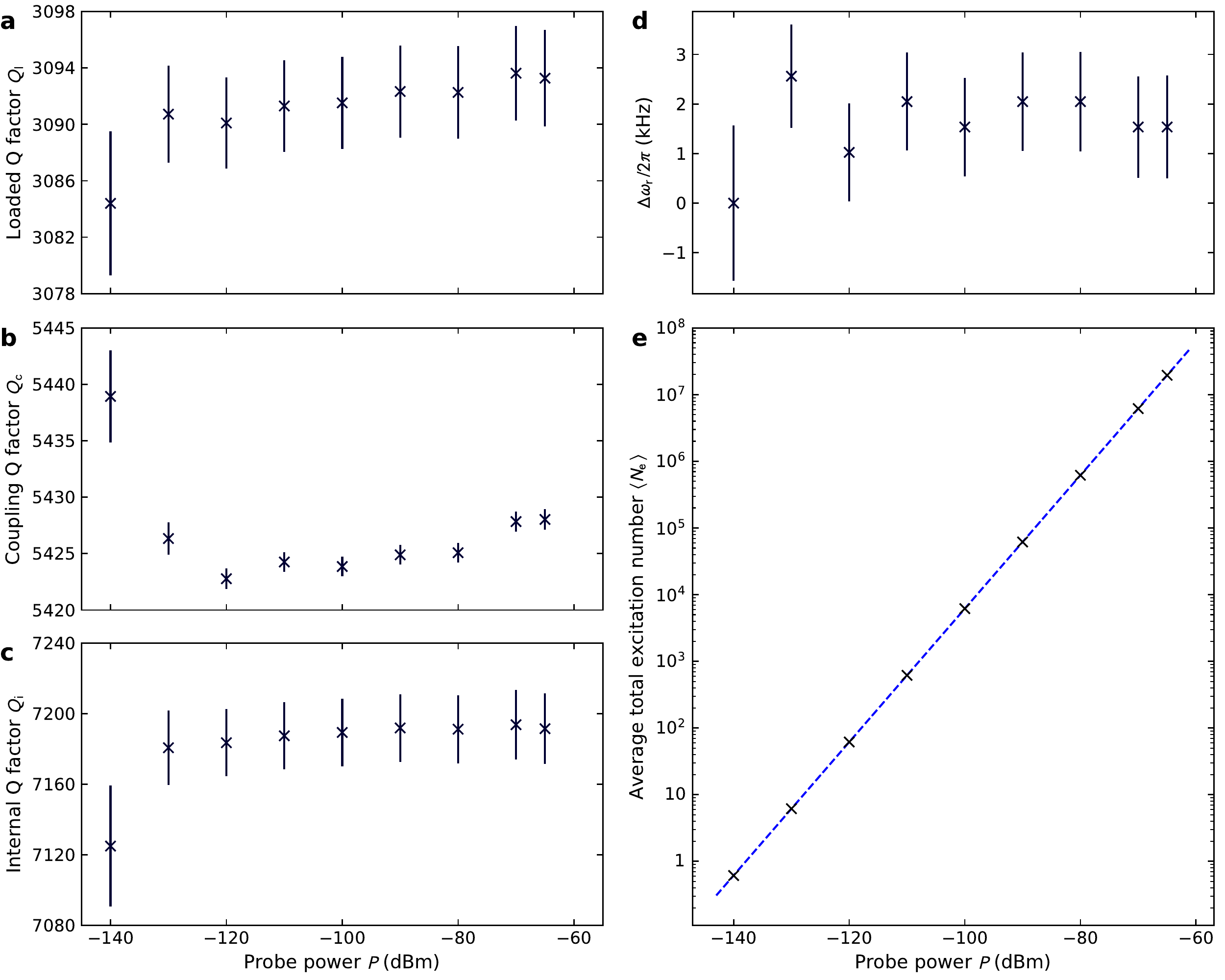}
\caption{
(a-c) Loaded, coupling, and internal quality factors of the cavity resonance against probe power. The data was taken at $T=\SI{55}{\milli\kelvin}$ with zero current applied to the magnetic coil and does not show a power dependent behavior. (d) Shift of the fitted cavity frequencies $\Delta\omega/2\uppi=\left[\omega_\mathrm{r}\left(P\right)-\omega_\mathrm{r}\left(P=\SI{-140}{\deci\bel\metre} \right)\right]/2\uppi$ with compared to the measurement at lowest probe power at zero current. Similar as with the quality factors, the cavity frequency does not show a power dependence. (e) Calculated average photon number in cavity against probe power. The fit shows a linear dependence of the photon number calculated with Eq.\@ (\ref{app:eq:photonnumber}) to the input power. Note that this plot features a log-log scale, making the fit linear again. The errors on the average photon number are estimated to be smaller than $\SI{0.35}{\percent}$ and are not visible on this plot.
}
\label{app:fig:excitationnumber}
\end{figure*}

Using the cavity's resonance frequency, quality factors, and the input power $P_\mathrm{in}$ we estimate the total number of magnon and photon excitation within the cavity $\left<N_\mathrm{e}\right>$ in units of $\hbar\omega_\mathrm{r}$,
\begin{equation}
	\left<N_\mathrm{e}\right> = 4\frac{Q_\mathrm{l}^2}{Q_\mathrm{c}}\frac{1}{\hbar\omega_\mathrm{r}^2} \cdot P_\mathrm{in}.
	\label{app:eq:photonnumber}
\end{equation}
Note that the input power $P_\mathrm{in}$ is in units of watts and not to be confused with the probe power (level) $P$ in units of $\mathrm{dBm}$. For the strongly coupled system the excitation energy at matching frequencies is stored in equal parts in photons and magnons, $\left<n\right>=\left<m\right>=\frac{\left<N_\mathrm{e}\right>}{2}$. We measure the reflection signal of the cavity resonance at $\SI{55}{\milli\kelvin}$ at probe powers between $-140\,\mathrm{dBm}$ and $-65\,\mathrm{dBm}$ at zero current applied to the magnetic coil. The complex data is then fitted using a circle fit algorithm \cite{Probst_2015} to determine the power dependent quality factors (coupling quality factor $Q_\mathrm{c}=\frac{\omega_\mathrm{r}}{2\kappa_\mathrm{c}}$, internal quality factor $Q_\mathrm{i}=\frac{\omega_\mathrm{r}}{2\kappa_\mathrm{i}}$, and loaded quality factor $Q_\mathrm{l}=\left(1/Q_\mathrm{i}+1/Q_\mathrm{c}\right)^{-1}$) and resonance frequencies as shown in Fig.\@ \ref{app:fig:excitationnumber} (a-d). Besides an initial shift of the quality factors of less than $\SI{1}{\percent}$ going from $-140\,\mathrm{dBm}$ to $-130\,\mathrm{dBm}$ of input power the quality factors show no power dependence, varying only in a range below $\SI{0.15}{\percent}$. The fitted (dressed) frequencies are shifted compared to the bare cavity frequency due to the residual magnetic field. From Eq.\@ (\ref{app:eq:bare_frequency}) we expect a zero-current dressed frequency of $\omega_\mathrm{r}^{I=0}=\SI[separate-uncertainty, multi-part-units=single]{5.239452(2)}{\giga\hertz}$. The circle fit gives a resonance frequency at the lowest power of $\omega_\mathrm{r}^\mathrm{CF}=\SI[separate-uncertainty, multi-part-units=single]{5.239474(2)}{\giga\hertz}$. 
We calculate the average total excitation for all probe powers using Eq.\@ (\ref{app:eq:photonnumber}) and fit a line to the logarithmic data (Fig.\@ \ref{app:fig:excitationnumber} (e)),
\begin{equation}
	\left<N_\mathrm{e}\right> = 62.046\cdot P_\mathrm{in}^{1.0003}\,\si{\per\femto\watt}.
	\label{app:eq:photonnumber_fit}
\end{equation}
The fit agrees well with the data and is used throughout all data evaluation to map probe powers $P$ to average excitation numbers. This results in average magnon excitation numbers at matching frequencies for our experiments between $0.31$ and $9.85\cdot 10^6$. 

\section{\label{app:sec:kappa_m}Extracting the internal magnon linewidth}
We extract the internal magnon linewidth $\kappa_\mathrm{m}$ by fitting the reflection amplitude $\left|\mathcal{S}_{11}(\omega)\right|$ using the input-output formalism \cite{QuantumOpticsBook}.
\begin{equation}
\begin{split}
	&\left|\mathcal{S}_{11} (\omega_\mathrm{p}, I)\right|=\\
	& \quad \quad \left|-1 +\frac{2\kappa_\mathrm{c}} {\mathrm{i} \left(\omega_\mathrm{r}-\omega_\mathrm{p}\right) +\kappa_\mathrm{l}+ \frac{g^2}{\mathrm{i}\left(\omega_\mathrm{m}\left(I\right)-\omega_\mathrm{p}\right)+\kappa_\mathrm{m}}}\right|, 
	\end{split}
	\label{app:eq:abs_S11}
\end{equation}
with the probe frequency $\omega_\mathrm{p}$, the magnon frequency $\omega_\mathrm{m}$, and the loaded, coupling and magnon linewidths $\kappa_\mathrm{l}$, $\kappa_\mathrm{c}$, and $\kappa_\mathrm{m}$ (HWHM). 
Before fitting, we normalize the data by the current independent baseline similar to Ref. \cite{Boventer_2018}. We estimate the background value for each probe frequency by calculating a weighted average over all entries along the current axis, neglecting the areas around the dressed cavity resonances. The amplitude data is divided by this baseline to account for losses in the measurement setup. The normalized data together with the fit results of Eq.\@ (\ref{app:eq:bare_frequency}) and the circle fitted cavity resonance at zero current are then fitted to Eq.\@ (\ref{app:eq:abs_S11}) using the Python package lmfit \cite{lmfit}. 

\begin{figure}[htb]
\centering
\includegraphics[width=\columnwidth]{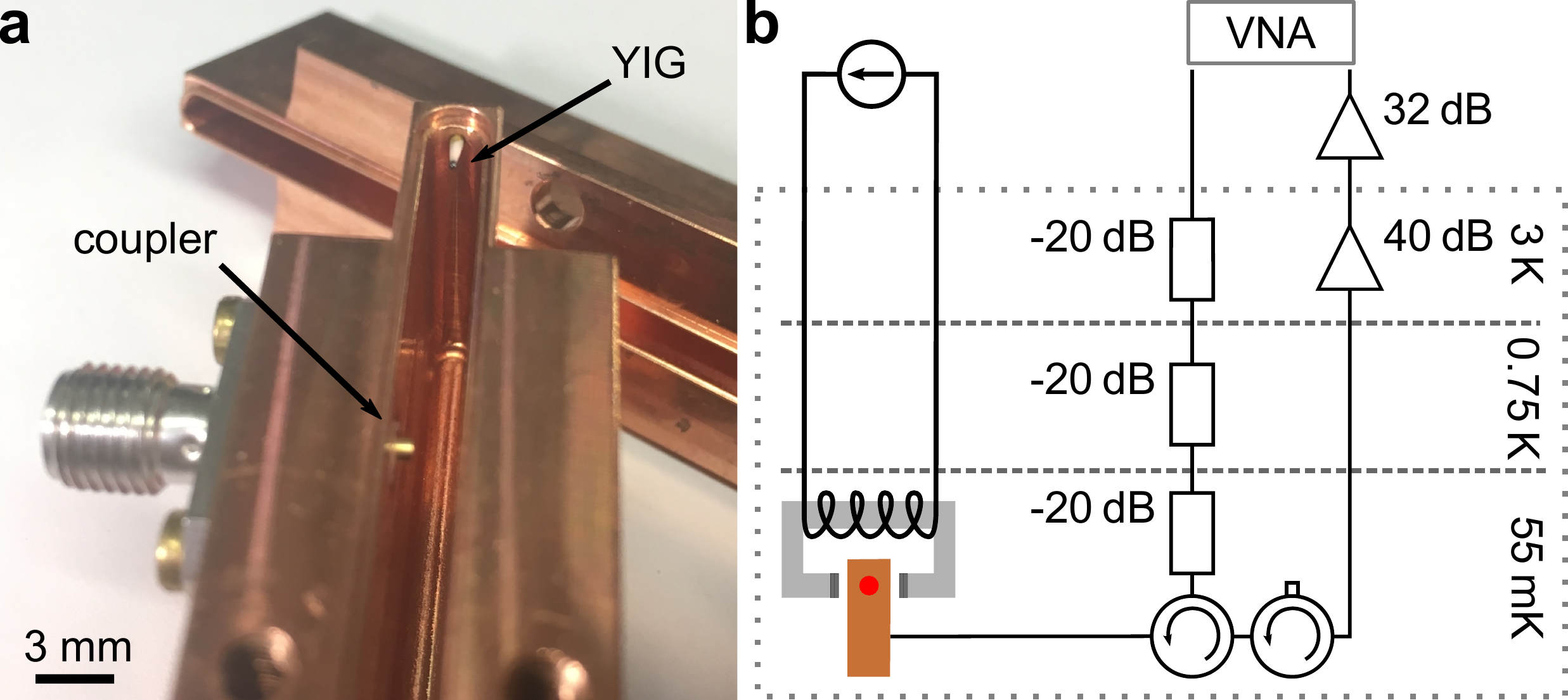}
\caption{
(a) Photograph of the sample in the cavity. The top half of the cavity resonator was removed and can be seen in the background.
(b) Schematic diagram of the experimental setup. The cavity holding the YIG sphere and the magnet providing the static field are mounted at the mixing chamber plate of a dry dilution refrigerator. The microwave input signal is attenuated to minimize thermal noise at the sample. The attenuation of the complete input line to the input port of the cavity is $\SI{-75}{\deci\bel}$ at the cavity resonance frequency. The output signal is amplified by a cryogenic amplifier operating at $\SI{3}{\kelvin}$ and an amplifier at room temperature. Two magnetically shielded microwave circulators protect the sample from amplifier noise.
}
\label{app:fig:experimental_setup}
\end{figure}

\section{\label{app:sec:TLS}Possible TLS origin}
The microscopic origins of TLSs is still unclear and part of ongoing research. Possible models include magnetic TLSs proposed with analog behavior to the electric dipolar coupled TLSs \cite{Continentino_1981_ssc, Continentino_1981, Continentino_1983, Continentino_1983_jpc} and measured in spin glasses by thermal conductivity, susceptibility and magnetization measuements at low temperatures \cite{Arzoumanian_1983, Wassermann_1984}. With amorphous YIG showing spin glass behavior \cite{Wesenberg_2017} it seems plausible to observe these effects in our crystalline YIG sample where in addition to the observed rare earth impurities \cite{Boventer_2018} we can assume structural crystal defects. This is based on materials with electric dipolar coupled TLSs, where TLSs appear largely in disorderd crystals but also in single crystals with smaller density \cite{Kleinman_1987}.

\begin{figure*}[htb]
\centering
\includegraphics[width=\textwidth]{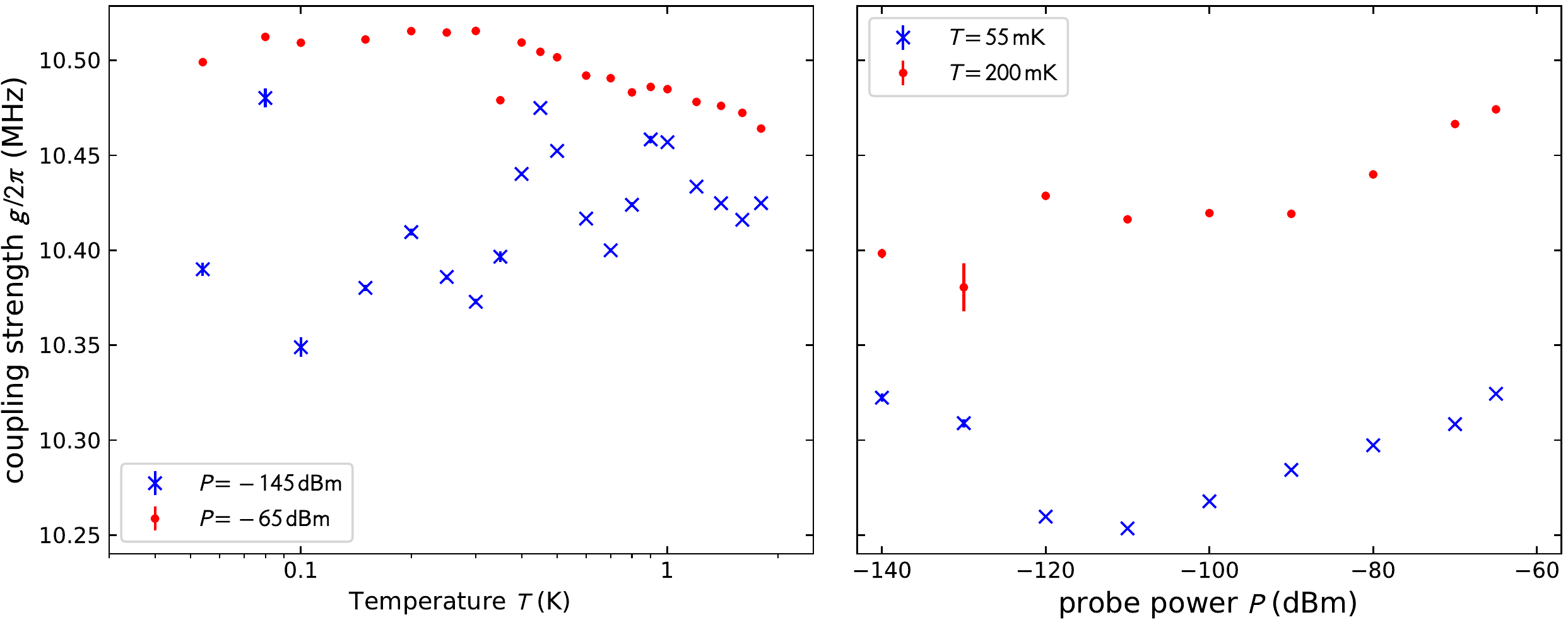}
\caption{
(a) Temperature and (b) power dependence of the coupling strength evaluated at the same conditions as Fig. (2) in the main text. We find a increase of the coupling strength of about $1\,\%$ going to higher powers that decreases at higher temperatures. This indicates an increase in participating spins on the order of $2\,\%$.
}
\label{app:fig:coupling_strength}
\end{figure*}

Another possibility could be surface spins leading to strong damping that were observed as an important loss mechanism in cQED experiments \cite{deGraaf_2017, deGraaf_2018}. We evaluated the coupling strength to find a power or temperature dependence on the participating number of spins, see Fig. \ref{app:fig:coupling_strength}. We find an increase in the coupling strength of about $1\,\%$ at the saturation conditions for the TLSs. With $g\propto \sqrt{N}$ this translates to an increase in the number of participating spins of the order of $2\,\%$, e.g. due to the increased participation of now environmentally decoupled surface spins. This should not be enough to explain the decrease in $\kappa_\mathrm{m}$ by a factor of 2.

A loss mechanism by magnon-phonon coupling and subsequent phonon losses due to TLS coupling can be neglected since for $\mathbf{k}=0$ magnons in YIG these magnon losses are proposed to be much smaller than the Gilbert damping \cite{Streib_2019}.

\end{document}